\documentclass[a4paper,onecolumn,twoside,fleqn,10pt]{article}
\usepackage{amsmath}
\usepackage{authblk}
\usepackage[english]{babel}
\usepackage{bm}
\usepackage[font={color=blue},figurename=Fig.,labelfont={it}]{caption}
\usepackage{cite}
\usepackage{enumitem}   
\usepackage{fchicago} 
\usepackage{float}
\usepackage[left=2cm,right=2cm,top=2cm,bottom=2cm]{geometry}
\usepackage{graphicx}
\usepackage{hyperref}
\usepackage{indentfirst}
\usepackage[utf8]{inputenc}
\usepackage[switch,columnwise]{lineno}
\usepackage{lipsum} 
\usepackage{pdfpages}
\usepackage{rotating}
\usepackage{setspace} 
\usepackage{subcaption}
\usepackage[varg]{txfonts}
\usepackage{wasysym}

\definecolor{blue}{RGB}{0,0,255}
\hypersetup{
    colorlinks=true,                          
    linkcolor=blue, 
    citecolor=blue, 
    urlcolor=blue,
    linktoc=all
}

\newcommand{\eg}{\textit{eg. }}
\newcommand{\cf}{\textit{cf. }}
\newcommand{\ie}{\textit{ie }}

\newcommand{\redtext}[1]{\textcolor{red}{#1}}

\usepackage{fancyhdr}
\begin{document}

\title{\huge On the effect of the luni-solar gravitational attraction on trees \\}


\author[1]{Le Mouël Jean-Louis}
\author[2]{Gibert Dominique}
\author[3]{Boulé Jean-Baptiste}
\author[4]{Zuddas Pierpaolo}
\author[5]{de Bremond d'Ars Jean}
\author[1]{Courtillot Vincent}
\author[3]{Lopes Fernando}
\author[6]{Gèze Marc}
\author[4]{Maineult Alexis}

\affil[1]{\small UMR7154, Universit\'e de Paris, Paris, France}
\affil[2]{\small UMR5276, Laboratoire de g\'eologie de Lyon Terre, plan\`etes et environnement (LGL-TPE), Lyon, France}
\affil[3]{\small UMR7196, Museum National d'Histoire Naturelle, Paris, France}
\affil[4]{\small UMR7619, Sorbonne Universit\'e, Paris, France}
\affil[5]{\small UMR6118, Universit\'e de Rennes, G\'eosciences Rennes, Rennes, France}
\affil[6]{\small UMR7245, Museum National d'Histoire Naturelle, Paris,France}

\date{\today}
\maketitle

\begin{abstract}

In this study, we revisit an old series of continuous electrical measurements made in 2003 on a standing Poplar tree in the village of Remungol, France (\cf \shortciteN{Gibert2006}), through the lens of a more refined signal analysis (Singular Spectrum Analysis). These measurements initially revealed the presence of a diurnal electrical signal, even in winter, that appeared to be closely correlated with sap flow. This observation led to the hypothesis that the electrical signal was the result of a simple electrokinetic phenomenon, as traditionally understood in geophysics. In the 2003 experiment, electrodes were placed in the roots, trunk and branches of the poplar. Our current analysis shows that the pseudo-diurnal cycle is not the only earth-tide identifiable during the year-long recording. Over 80\% of the variability in the signals is made up of seven of the major lunar-solar tidal cycles. An order of magnitude calculation gives an electrokinetic coupling coefficient similar to that of rocks, which supports the following hypothesis: sap flow variations are largely driven by first-order variations in the gravitational forces of the Moon and Sun, in the same way that these forces influence the movement of all fluids on the Earth's surface. Thus, sap circulation in a tree would not be due to osmotic pressure, evapotranspiration or capillarity.  

\end{abstract}

\section{\label{sec01} Introduction} 
	Trees, like the majority of plants on our planet's surface, can be broadly schematized as having a root system connected to branches through a trunk. Channels, such as phloem or xylem, originating from the rhizoderm and extending to the leaves, play a crucial role in facilitating proper irrigation and supplying nutrients for the plant's growth. The fluid, varying in complexity based on its location within the plant, functions as its circulatory system and is commonly referred to as sap. A significant and ongoing debate revolves around the question of how this sap moves within the plant. In the literature, three main hypotheses are typically discussed to explain this movement. The first one involves capillarity (\eg \shortciteNP{Pickard1981,Tyree1990,Kang2003,Brown2013,Li2021}), the second revolves around osmotic pressure (\eg \shortciteNP{Munns1984,Zimmermann1994,Scheenen2007,Jensen2016,Munns2020}) and the third, frequently debated hypothesis is that of evapotranspiration (\eg \shortciteNP{Granier1987,Smith1996,Poyatos2016,Poyatos2020}). This idea originates from climatologist  \shortciteN{Thornthwaite1948}, who sought to identify various geophysical phenomena contributing to the transfer of liquid water from the Earth's surface to the atmosphere, ultimately classifying different climates. Plant evapotranspiration is recognized as one of these phenomena, alongside snow sublimation and the evaporation of free water.

	In their study of electrical signals measured on a standing poplar over the course of a year, \shortciteN{Gibert2006} demonstrated two important findings. Firstly, the recorded electrokinetic signals were associated with the sap flow in the tree (estimated by Ganier probes), and secondly, a diurnal oscillation, albeit modest, persisted even during the winter season. 
	
	For several decades in geophysics, it has been established and observed that changes in velocity in an established flow or the addition of a saline front lead to measurable variations in electrical potential. This electrokinetic phenomenon is commonly referred to as spontaneous potential (SP, \eg \shortciteNP{Maineult2005,Maineult2008,Allegre2014}). In the natural environment, diurnal oscillations are not solely attributable to thermal fluctuations resulting from the alternation of day and night. Instead, they are generally linked to terrestrial gravitational tides caused by the combined influences of celestial bodies such as the Moon and the Sun  (\cf \shortciteNP{Laplace1799}). Our planet experiences disruptions during its revolution and rotation, with the Moon and the Sun playing pivotal roles. These celestial interactions lead to phenomena like the precession of the equinoxes, which occurs over a period of approximately 26,000 years (for further details, see, \eg \shortciteNP{Milankovic1920,Lopes2022a}). In a broader context, mobile masses on the Earth's surface oscillate under the influence of these luni-solar forces, with coastal tides being a primary manifestation. These forces also impact and vertically oscillate water confined in deep aquifers (\eg \shortciteNP{Narasimhan1984,Rojstaczer1990,Li2002,Dumont2020,Lordi2022}), potentially influencing processes such as water-rock interactions (\eg \shortciteNP{Brantley2008,Scislewski2010,Zuddas2024}). Consequently, these interactions may play a role in providing nutrients to plants (\eg \shortciteNP{Barbera2023}).
	
	We have demonstrated, over long periods exceeding 200 years, how the global series of volcanic eruptions activity (\cf \shortciteNP{LeMouel2023}), the average sea levels evolution (\cf \shortciteNP{Courtillot2022}) and the dynamics of the trees population within a Tibetan juniper forest (\cf \shortciteNP{Courtillot2023}), could be decomposed into a series of cyclical patterns, all linked to the gravitational forces, this time originating from the Jovian planets. The first two phenomena being related to vertical movements of fluids. As previously explained, tidal phenomena exhibit a multi-scale nature in time, spanning from a few hours to thousands of years, and are marked by extensive spatial dimensions. In this presented study, utilizing both historical observations and contemporary data, our objective is to extend the conclusions drawn by \shortciteN{Gibert2006}) and provide further clarity on the driving mechanisms reflected in the electrical signals of trees.

	After briefly recalling the electrical acquisition system used on the poplar tree in Remungol in 2003 and presenting some of the data (section \ref{sec02}), we will revisit the phenomenon of earth-tides, which will be introduced theoretically (section \ref{sec03}). In section (\ref{sec04}) we will present the results of the analysis, with a detailed discussion in the appendix (\Alph{section}), and propose a mechanism before concluding (section \ref{sec05}).

\section{\label{sec02} Back to the Remungol standing Poplar experiment}
\begin{figure}[H]
		\centering{\includegraphics[width=\columnwidth]{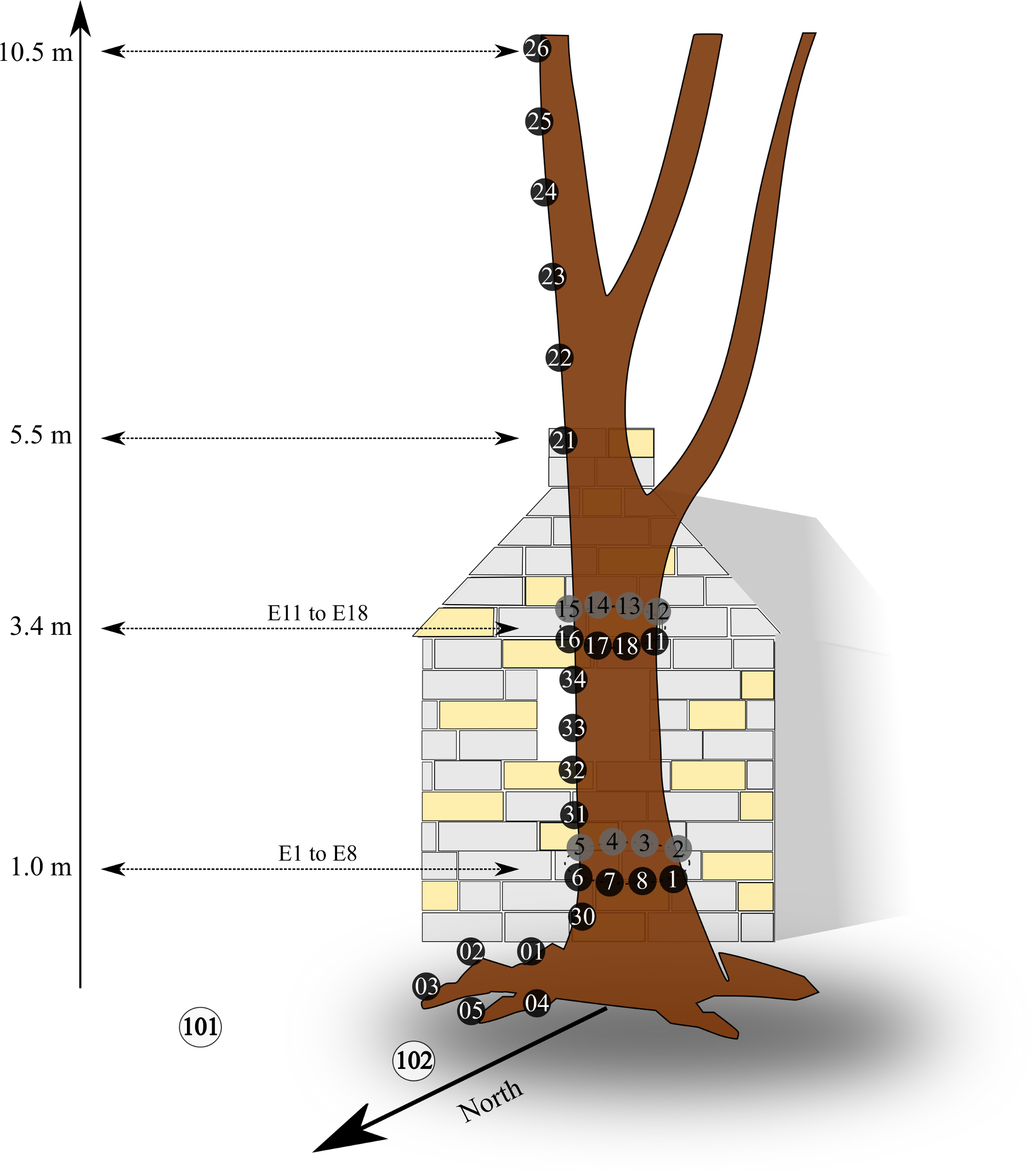}} 	
     	\caption{Schematic representation of the electrical acquisition protocol conducted on the Remungol poplar tree in 2003. Measurements were taken all over the tree: at the roots (electrodes 01 to 05), in two circles around the trunk, the first at 1 metre above the ground (electrodes 1 to 8), the second at 3.4 metres above the ground (electrodes 11 to 18), on a generatrix on the north-facing side (electrodes 30 to 35) and finally on a branch starting at 5.5 metres above the ground (electrodes 21 to 26).}
		\label{Fig:01} 	
\end{figure}

\begin{figure}[H]
		\centering{\includegraphics[width=\columnwidth]{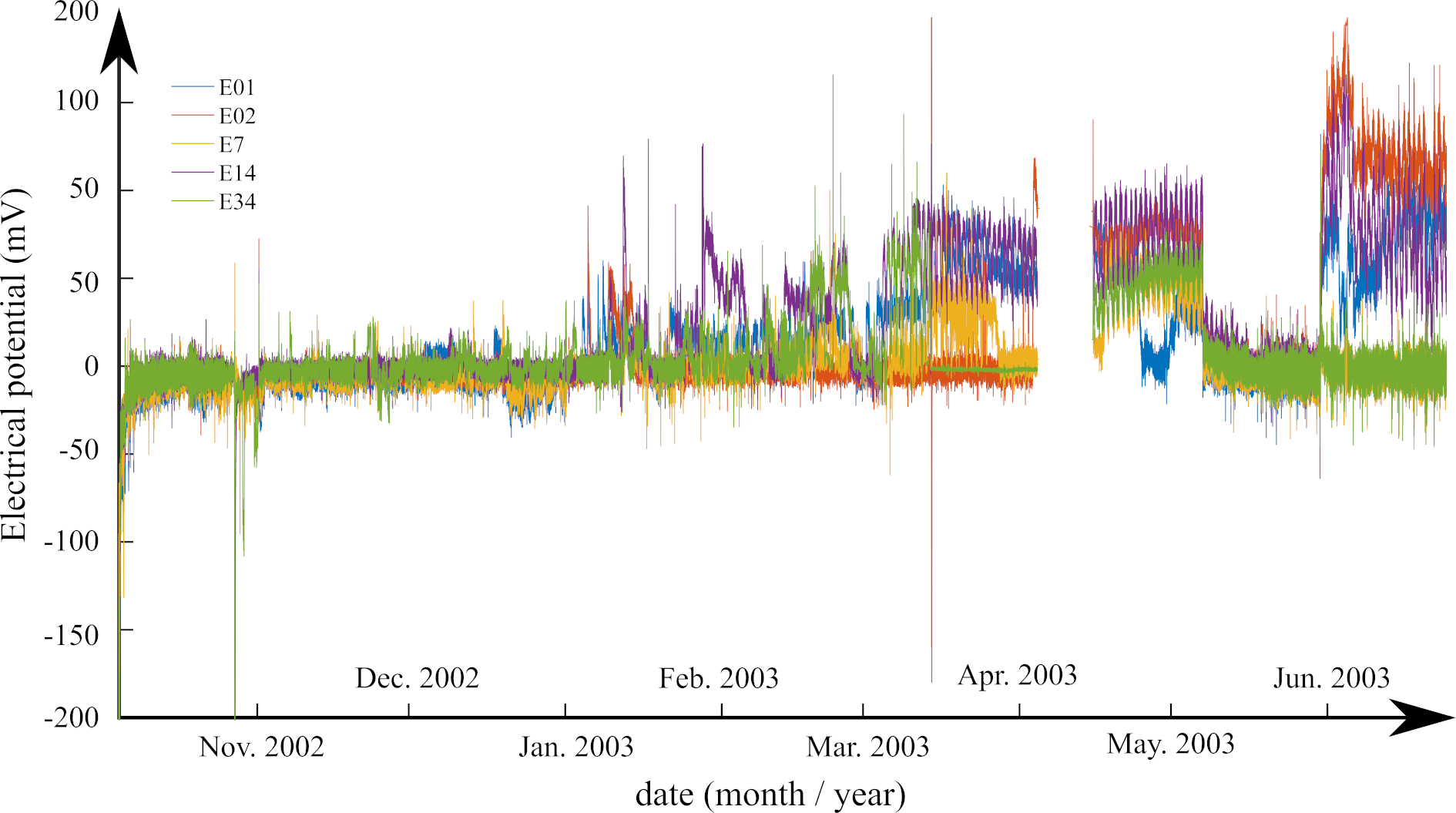}} 	
     	\caption{Illustration of some recorded potentials on both the roots and trunk of the poplar tree. One can observe the non-stationary nature of the potentials as well as data gaps in April 2003}
		\label{Fig:02} 	
\end{figure}

	The experiments carried out by \shortciteN{Gibert2006} on a poplar tree (\cf Figure \ref{Fig:01} for the acquisition protocol diagram) was conducted in two steps. The first phase ran from 13 September 2002 to 29 October 2003, followed by a second phase from 25 November 2003 to 21 May 2005. The 2006 paper analyses data from the second campaign. However, since this phase has significant periods without records due to the upgrade of the acquisition system, we will focus here on the first dataset, which has only a small gap of 9 days in April 2003 (\cf Figure \ref{Fig:02}). The aim is to study the evolution of the envelopes of the different pseudo-periods recorded over the course of a year. We briefly recall, and present in Figure \ref{Fig:01}, the acquisition protocol that was used in 2002-2003 for the Remungol poplar tree experiment. On that occasion, 32 stainless steel electrodes were inserted into the tree and 2 non-polarizable electrodes (\shortciteNP{Petiau1980}) were buried in the soil,
\begin{enumerate}[label=(\roman*)]
	\item 2 non-polarizable Petiau \shortcite{Petiau1980} electrodes in the ground (number 101 and 102),
	\item 5 stainless steel electrodes on the visible roots of the poplar tree (number 01 to 04),
	\item 2 stainless steel sets of electrode crowns, one positioned at a height of 1 meter from the ground and surrounding the trunk (number 1 to 8) and the other positioned at 3.4 meters from the ground (number 11 to 18),
	\item 2 stainless steel sets of 5 electrodes each on the north-facing side of the trunk, one between 0.9 and 3.4 meters (number 30 to 34) the other, from 5.5 meters to 10.5 meters (number 21 to 26)
\end{enumerate}	
\newpage
	
	The measuring instrument employed is a Keithley 2701 digital multimeter with an input impedance greater than 100 M$\Omega$, equipped with a relay matrix having 40 measurement channels controlled by acquisition software. Electric potential measurements are taken at all electrodes at a sampling interval of 1 minute. A UT time base was used, synchronized in real-time with the Frankfurt atomic clock. Both the computer and the multimeter are powered by a backup generator, mitigating interruptions due to brief electrical power line failures.
	
	In Figure \ref{Fig:02}, we present an example of recorded data from both the roots, the trunk, and the branches during the period from October 1, 2002, to the end of June 2003. On the x-axis, which represents time, each tick corresponds to the first day of each month.

Subsequently, we will present and analyze this time range, which is the longest with minimal gaps, covering transitions through each of the seasons. As illustrated in this figure, the electrical potentials, both in the roots, trunk, and branches, can vary by several hundred millivolts over the 320 days depicted. An activity (early) appears to commence in January 2023, reaching its peak during the summer of the same year. Clearly, the signals are not stationary, neither in the strict sense nor in the broader sense. Peaks of polarization are observed at times, and there is a gap in the data. These symptoms led us to choose Singular Spectrum Analysis as the method for analysis and extraction, which we will discuss later (\cf Appendix A).	Henceforth, when referring to the data from  \shortciteN{Gibert2006}, we will use the acronym \textbf{RP} (Remungol Poplar).

\section{\label{sec03} The Earth-tides: A Theoretical Overview}
	Earth tides are small (low amplitude) phenomena acting on the Earth by the Newtonian gravitational field of celestial bodies (primarily the Moon and Sun). Since \shortciteN{Laplace1799} we have been able to calculate the influence of the gravitational potential of any celestial body, such as the Moon, on the oceans. This tidal potential ($\mathcal{W}$) is given by the following relation,
\begin{equation}
	\mathcal{W} = \dfrac{\mathcal{G}M}{D} \sum_{n=2}^{\infty} (\dfrac{R}{D})^{n}\mathcal{P}_{n} (\cos z).
	\label{eq:hs01}
\end{equation}

	In equation (\ref{eq:hs01}) we find the gravitational constant ($\mathcal{G}$), the mass of the Moon ($M$), the radius of the Earth ($R$) and the Earth-Moon distance ($D$). $\mathcal{P}_{n}(\cos z)$ is the \shortciteN{Legendre1785} polynomial in the cosine of the angle $z$, which is the angle between the centre of the Earth, the geographical point on its surface where the potential acts, and the position of the Moon in the celestial sphere.	For a better understanding of this expression, called spherical harmonic decomposition, we invite the reader to refer to \shortciteN{LeMouel2023b}, which describes exactly how it is obtained for both the electric and magnetic fields. The scalar decomposition remains the same for the gravitational field. Since the celestial bodies in the solar system are at considerable distances $D$ from the Earth's radius $R$, we generally restrict ourselves to the first term in the series ($\mathcal{W}_{2}$) for the tidal potential. In this case, as mentioned above, only the Moon and the Sun have a major influence on the tidal potential, with the Moon contributing a factor of 2 and the Sun a factor of 1. The effects of these two celestial bodies combine in a complex way, taking into account their positions relative to the Earth. The resulting effect is greatest when the Moon and Sun are in conjunction or opposition, and especially during eclipses.
	
	The ratios of the spatial dimensions involved in equation (\ref{eq:hs01}) are such that only the Legendre polynomials $\mathcal{P}_{n} (\cos z)$remain as variables in the present problem. To fully evaluate them, it is necessary to take into account not only the angle $z$, but also the right ascension ($H$) and the declination ($\delta$) of the celestial body (the Moon or the Sun), as well as the geographical coordinates (latitude $\theta$ and longitude $\varphi$) of the point where the tide occurs,
\begin{equation}
	\mathcal{P}_{n} (\cos z) = \mathcal{P}_{n} (\sin \theta).\mathcal{P}_{n} (\sin \delta) + \sum_{m=1}^{n} \mathcal{P}_{n}^{m}(\sin \theta).\mathcal{P}_{n}^{m}(\sin \delta).\cos m (H-\varphi),
	\label{eq:hs02}
\end{equation}	

	with the particular case, $n=1$: $\cos z = \sin \theta . \sin \delta + \cos \theta . \cos \delta . \cos (H-\varphi)$. As mentioned above, the tidal potential is the first term in the decomposition of (\ref{eq:hs01}). Thus, up to a common factor, the tidal potential $\mathcal{W}_{2}$  is decomposed into three terms,
\begin{itemize}
	\item zonal: $\mathcal{W}_{2,0} = \mathcal{P}_{2} (\sin \theta).\mathcal{P}_{2} (\sin \delta)$,
	\item tesseral: $\mathcal{W}_{2,1} = \mathcal{P}_{2}^{1} (\sin \theta).\mathcal{P}_{2}^{1}(\sin \delta) .\cos(H-\varphi)$,
	\item sectoral: $\mathcal{W}_{2,2} = \mathcal{P}_{2}^{2} (\sin \theta).\mathcal{P}_{2}^{2}(\sin \delta) .\cos2(H-\varphi)$.
\end{itemize}

\begin{figure}[H]
		\centering{\includegraphics[width=0.5\columnwidth]{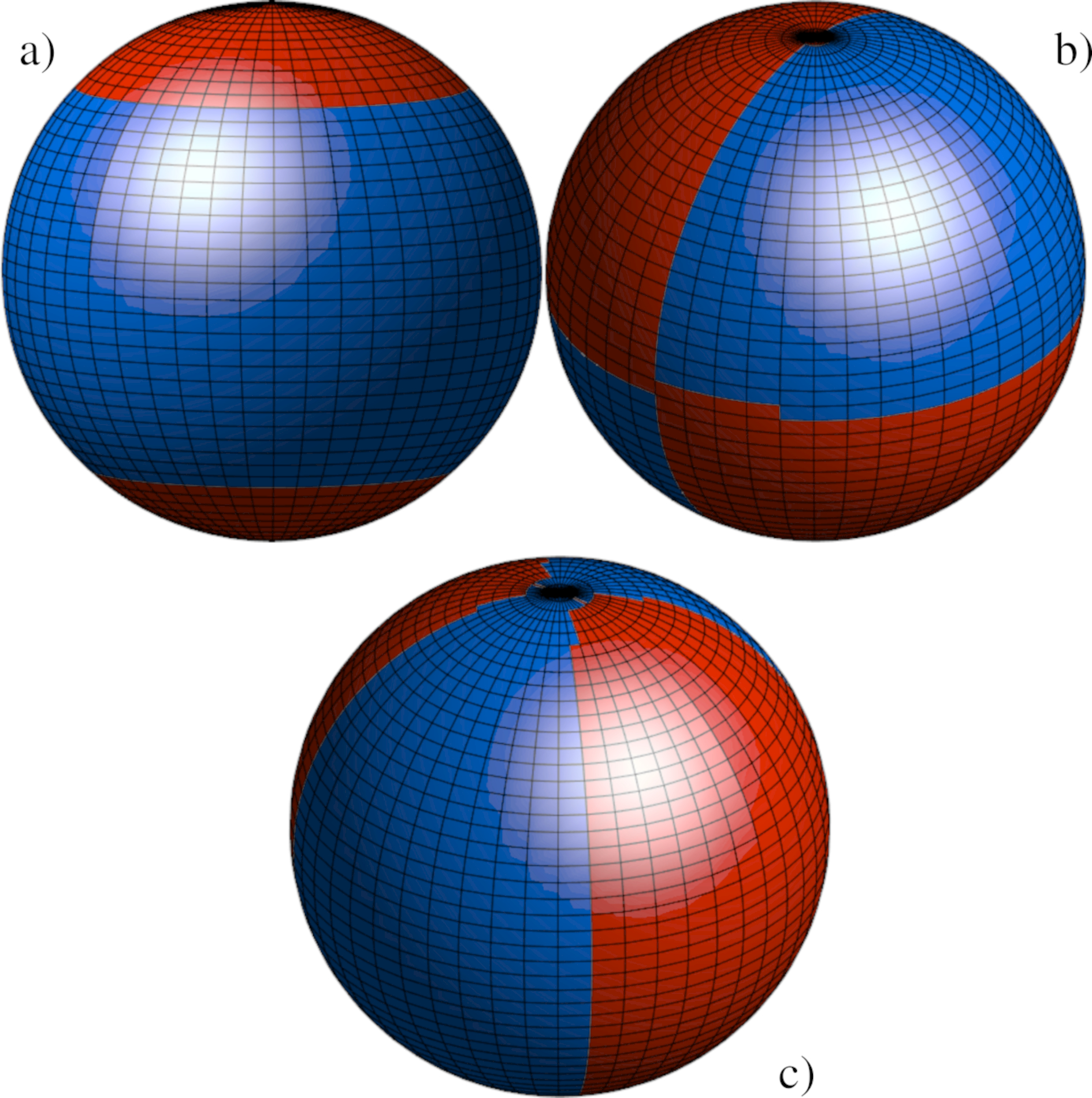}} 	
     	\caption{Distribution of the tidal potential. a) the zonal term, b) the tesseral term, and c) the sectorial term.}
		\label{Fig:earthtides} 	
\end{figure}

	By applying these last three relations to the combined ephemerides of the Moon and Sun, it is possible to calculate the full set of lunar-solar tides, both in terms of their periods and their intensities (\ie their accelerations). We have listed them in Table \ref{table01}. To better understand the physical meaning of these terms, they can be summarised as follows,
	
\begin{itemize}
	 \item  Zonal terms are symmetrical about the Earth's axis of rotation (the polar axis). They do not vary with longitude, only with latitude (\eg Figure \ref{Fig:earthtides}a).
	 \item  Tesseral terms vary with both latitude and longitude. They represent more complex tidal variations and are influenced by the relative position of the Earth and the celestial body (here the Sun) in both latitude and longitude (\eg Figure \ref{Fig:earthtides}b). 
	  \item  Sectoral terms vary mainly with longitude and are less dependent on latitude (\eg Figure \ref{Fig:earthtides}c).
\end{itemize}
	
	For example, the P1 Earth tide, the 7th potential found in Table \ref{table01}, which is a diurnal wave, the main Solar wave, has components that vary with both longitude and latitude, corresponding to a tesseral term in the classification of tidal waves.
	
\begin{table}[H]
\centering
\begin{tabular}{ p{2cm}|p{4cm}|p{2cm}|p{4cm}}
 \multicolumn{3}{c}{} \\
 \hline
  \hline
 Names  & Periodicities   & Magnitude ($cm^{2}/s^{2}$) & Origin \\
 \hline
 \ & \ \\
S$_{Sa}$   & 182.621 days  & 1909&  solar declination \\
M$_{m}$    & 27.555  days  & 2163&  lunar ellipticity   \\
M$_{f}$    & 13.661  days  & 4099&  lunar declination  \\
 \hline
Q$_{1}$   & 26h 53m  &1891&  lunar ellipticity   \\
O$_{1}$   & 25h 49m  &9876&  main lunar wave  \\
M$_{1}$   & 24h 50m  &1229&  lunar ellipticity   \\
P$_{1}$   & 24h 04m  &4600&  main solar wave  \\
K$_{1}$   & 23h 56m  &13902& solar and lunar declination (sideral day) \\
 \hline
Q$_{2}$   & 12h 39m  &4556&  lunar ellipticity   \\
M$_{2}$   & 12h 25m  &23798&  main lunar wave  \\
S$_{2}$   & 12h 00m  &11081&  main solar wave  \\
K$_{2}$   & 11h 58m  &3015& solar and lunar declination  \\ 
 \hline
\end{tabular}   
\caption{List of main luni-solar tides. In the first column, their associated names, in the second column, the period, and finally, in the last column, the celestial origin of the respective period.}
\label{table01}
\end{table}

\section{\label{sec04} Data analysis and model}
\begin{figure}[H]
		\centering{\includegraphics[width=1\columnwidth]{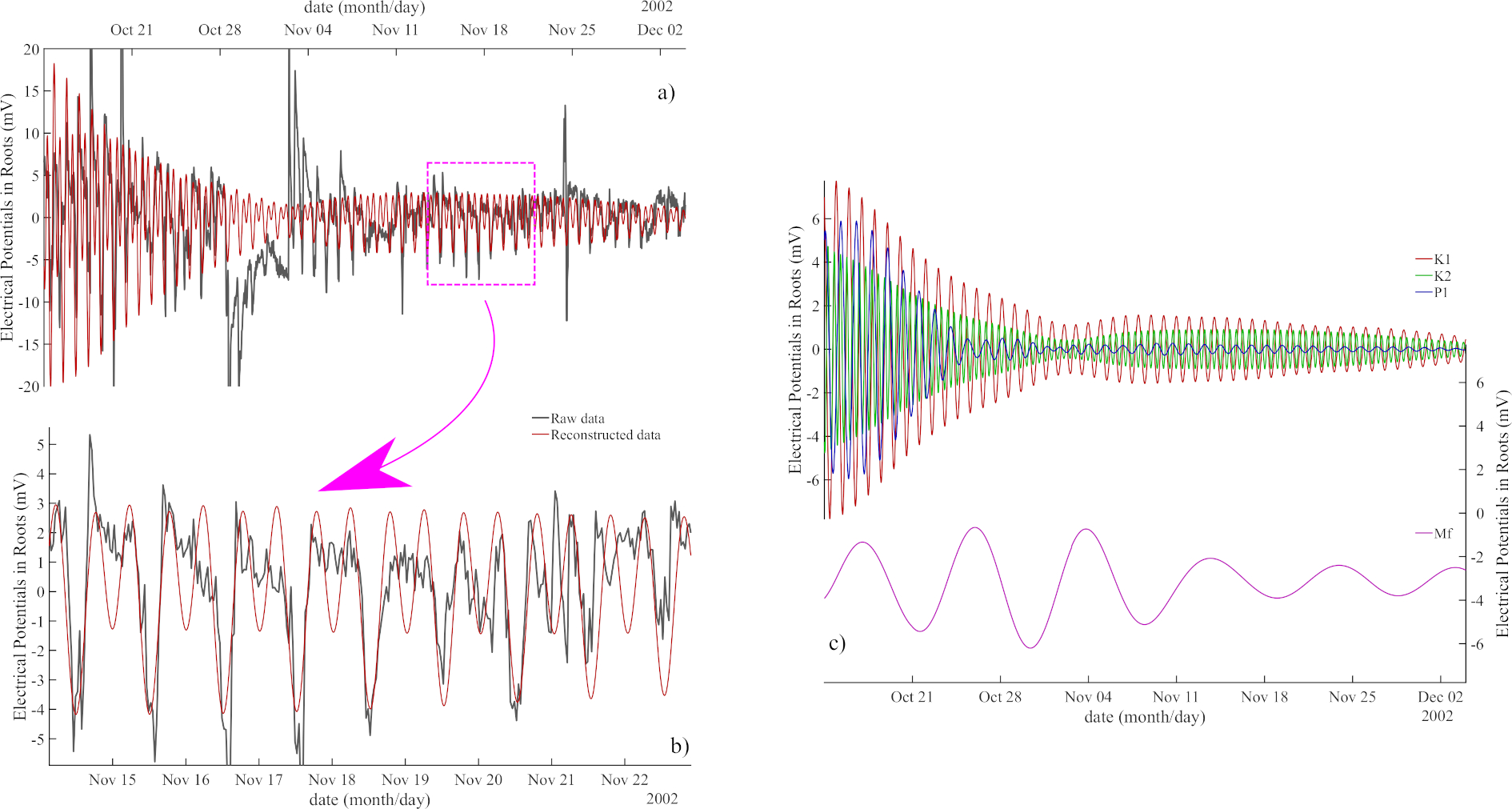}} 	
     	\caption{a) The raw electrical signal recorded by the E01 root electrode (gray curve). In red, the reconstruction obtained from the 5 extracted components.b) A zoom of a). c) The main earth tides extracted from the signal in a).}
		\label{Fig:E01rebuilt} 	
\end{figure}

	In a first step, we decompose the 2003 \textbf{RP} data using \textbf{SSA} for each of the 32 electrodes (see Figure \ref{Fig:02}. We observe that, in addition to diurnal and semi-diurnal oscillations previously identified in \shortciteN{Gibert2006}, more than 80\% of the data variability is carried by five major earth tides (see Table \ref{table01} and \shortciteNP{Coulomb1973,Ray2014}). An illustration of the latter point is given in Figure (\ref{Fig:E01rebuilt}). The electrical signal measured at the root by electrode E01 is shown in Figure \ref{Fig:E01rebuilt}a) between 13 October 2002 and 4 December 2002 for better readability. It was decomposed using \textbf{SSA} and the main components - K1, K2, P1 and Mf (see Table \ref{table01}) - are shown in Figure \ref{Fig:E01rebuilt}c). When these components are summed and superimposed on the original signal (see the red curve in Figure \ref{Fig:E01rebuilt}b), we observe a good reconstruction representing almost 80\% of the energy of the original signal.
	
\begin{figure}[H]
		\centering{\includegraphics[width=1.1\columnwidth]{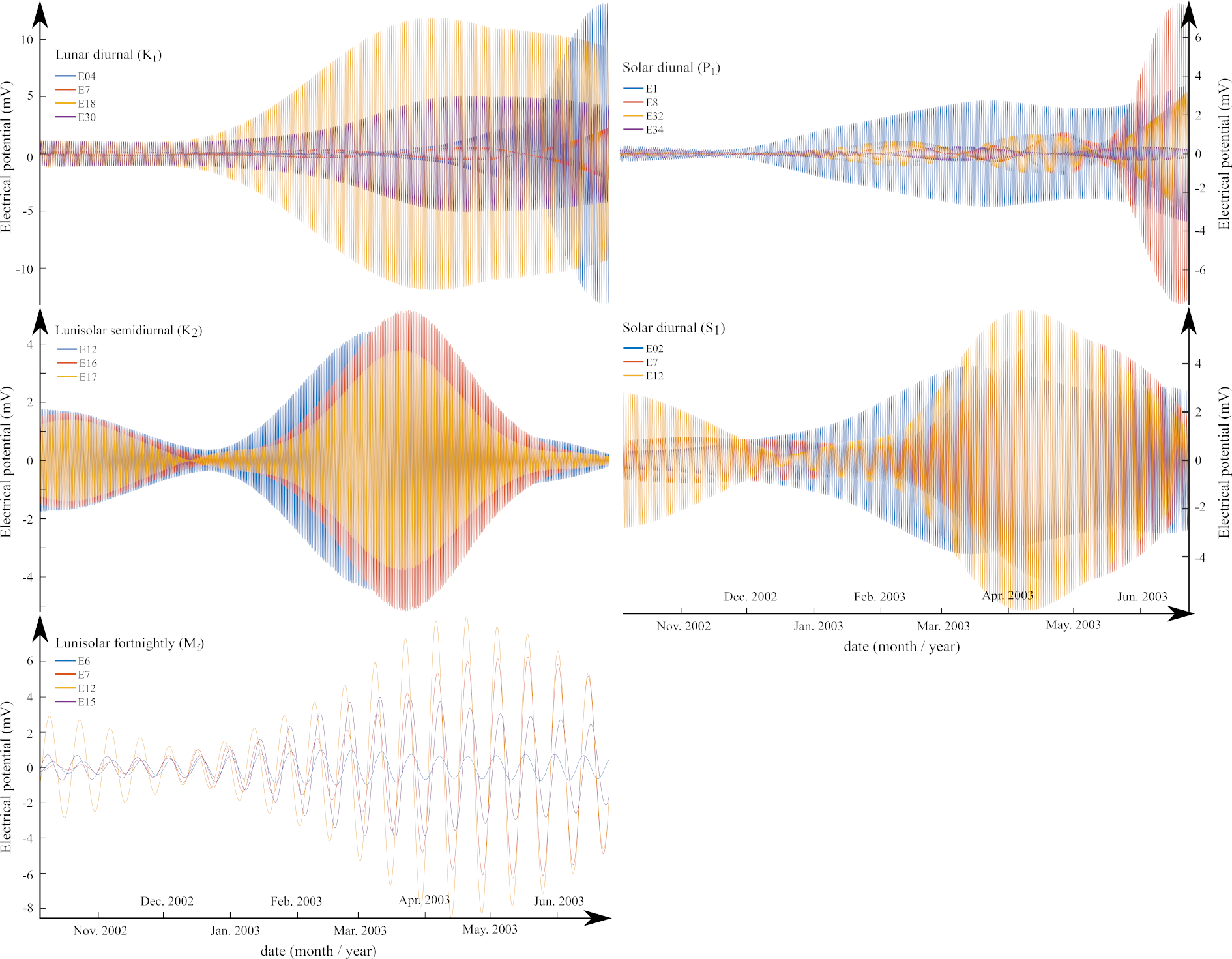}} 	
     	\caption{Top left: luni-solar tides K$_{1}$, extracted from electrodes E04, E7, E18, and E30.Top right: luni-solar tides P$_{1}$ extracted from electrodes E1, E8, E32, and E34. Middle left: luni-solar tides K$_{2}$ , extracted from electrodes E12, E16, E17. Middle right: luni-solar tides S$_{1}$ , extracted from electrodes E02, E7, E12. Bottom left: luni-solar tidess M$_{f}$, extracted by iSSA from electrodes E6, E7, E12,E13. }
		\label{Fig:05} 	
\end{figure}

	 For pedagogical reasons, we cannot overlay 32 curves on the same graph. The signals, once extracted by \textbf{SSA}, are regular enough for the calculation of a Fourier transform to make sense, in order to extract the nominal oscillation period of each of them. Examples of extraction results are shown in Figure \ref{Fig:05}.

\begin{figure}[H]
		\centering{\includegraphics[width=0.9\columnwidth]{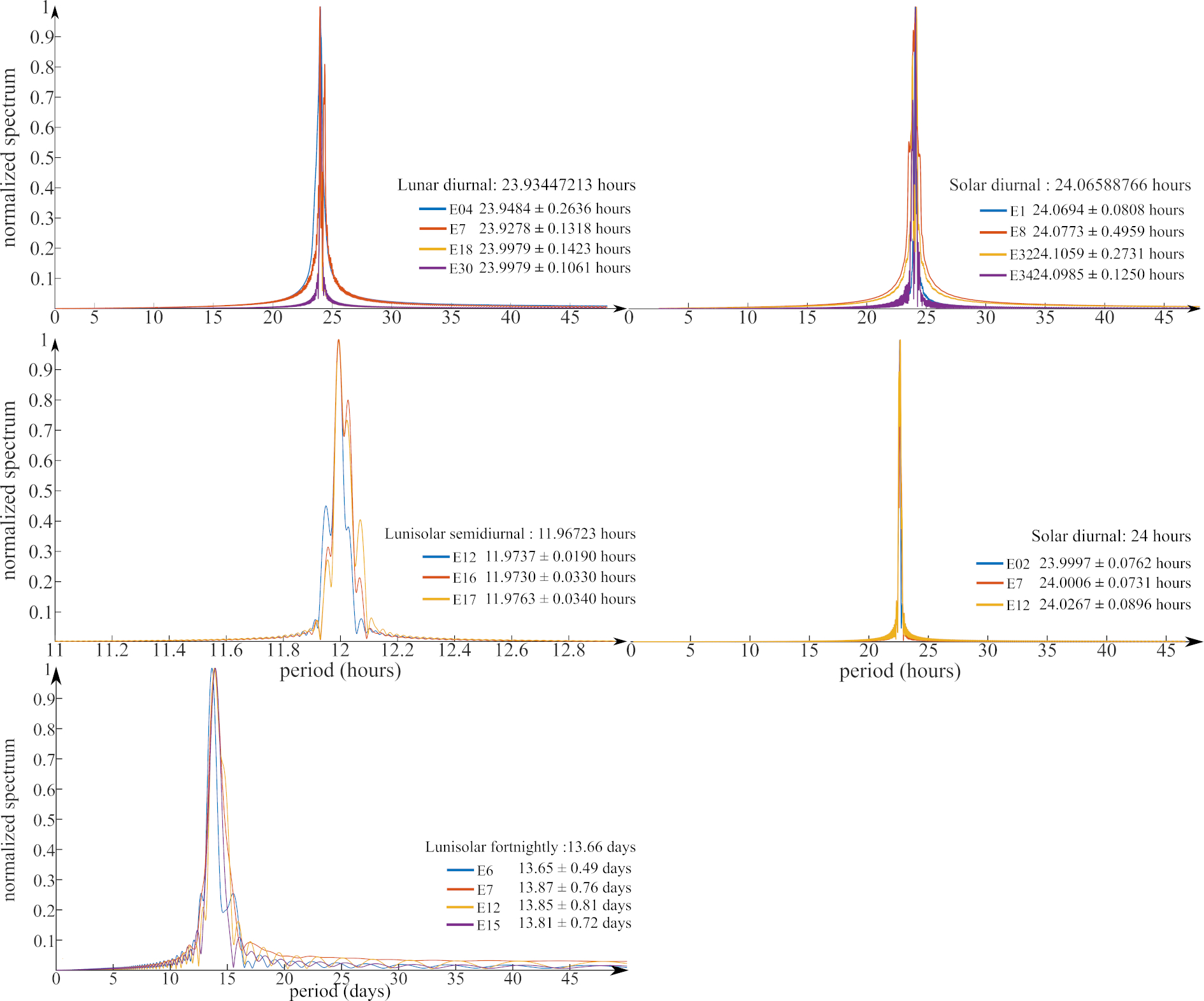}} 	
     	\caption{Fourier spectra corresponding to the signals presented in Figure \ref{Fig:05}. }
		\label{Fig:06} 	
\end{figure}

\paragraph{Presentation of some lunar-solar tides detected and extracted from the SP signals} \ \\
	In electrodes E1, E8, E32, and E34 (refer to our schematic in Figure \ref{Fig:01}), we have detected and extracted pseudo-cycles associated with the purely solar tide  P$_{1}$(refer to our Table \ref{table01}), which we present in Figure \ref{Fig:05} (top-right). In Figure \ref{Fig:06} (top-right), we show their respective Fourier spectra to determine the periodicities of these oscillations. These periodicities appear to be remarkably close, if not identical, to the ones obtained through calculations (refer to equations 01a and 01b), and we have indicated their theoretical values in the top-right corner of Figure 04b. Regarding the amplitudes, we observe a behavior that can only be attributed to the tree itself, as the variations in amplitudes are different at the same height ($\sim$1m) within the same crown (electrodes E1 and E8). We also note that these amplitudes evolve and seem to anticipate both solstices by approximately one month in all cases.\\
	
	We change the set of four electrodes for Figures \ref{Fig:05} (top-left) and \ref{Fig:06} (top-left) to demonstrate the generality of our findings. The second component we have detected and extracted is a purely lunar tide, the K$_1$  tide. Once again, as shown in the period spectra in Figure \ref{Fig:06} (top-left), the values of these periods are very close to the expected theoretical value of 23.93 hours. Similarly to what we observed for the S$_{1}$ tide, the amplitudes of the K$_1$ tide also appear to vary with the dates of solstices and equinoxes. The only difference here is that these variations are in phase with the mentioned dates. And as we observed for the previous tide, the potential increases from spring to reach its maximum in summer.

The others components, Figures \ref{Fig:05} (middle-left/right and bottom-left/right) and \ref{Fig:06} (middle-left/right and bottom-left/right), present in sequence the waveforms and period spectra of the following lunar-solar tides: S$_{1}$, the fortnightly M$_{f}$, and K$_{2}$. Similar to the examples we have described before, the waveforms of these lunar-solar tides shown in  Figures \ref{Fig:05} and \ref{Fig:06} are clearly modulated throughout the year, with a narrowing around the Winter solstice and a maximum amplitude shortly before the Summer solstice. These waveforms may not be in phase across different parts of the Remungol poplar tree (roots, trunk, branches), but in all cases, their calculated periods are very close to the expected values ($<$ 0.01$\%$). \\

For each of the 34 electrodes, it was possible to detect and extract between 7 luni-solar tides, which on average account for approximately 70\% of the raw variance for each signal. \\

\paragraph{A mechanism}\ \\
	Let us then envision briefly an electrokinetic mechanism. Suppose the tree contains channels, taken as regular cylinders reaching going up to the top of the tree, in which the sap circulates. This circulation is not yet well understood (see \shortciteNP{Zwieniecki2004,Windt2006,Zwieniecki2009,Vandegehuchte2013,Lopez2017,Li2021,Sakurai2021}).
	
	Let a channel of section $S$ of undetermined height. We will just make an order of magnitude computation of the pressure change in the channel due to the presence of a tidal forces of the Sun or (and) the Moon. Let us take the case of the Moon. $\mathcal{G}$ being the gravity of the Earth (in fact an acceleration in $m.s^{-2}$, and we neglect here the effect of the Earth’s rotation; $\mathcal{G} = 9.81 m.s^{-2}$ at the Earth’s surface). \\

	The vertical component the Moon tide, always an attraction, is on the Earth’s surface given by the following relation, 
\begin{equation}
	\Delta g_{r} = -\mathcal{G} \dfrac{m}{M}\dfrac{r^{3}}{R^{3}} (3.\cos^{2} \varphi -1) 
	\label{eq:01}
\end{equation}

$m$ is the mass of the Earth, $M$ the mass of the Moon, $r$ the mean radius, $R(t)$ the Earth-Moon distance, $\varphi$ the observation latitude. A similar relation gives the Solar tide, it is enough to replace the Earth-Moon distance by the Earth-Sun distance, and the mass of the Moon by that of the Sun.  Ignoring the latitude factor in \ref{eq:01}, and taking into account that $\dfrac{m}{M} = \dfrac{1}{80}, \dfrac{r}{R(t)} \sim \dfrac{r}{R} \approx 1.662*10^{-2}$ it comes $\Delta g_{r} \sim 0.566*10^{-6} m.s^{-2}$ . Let us consider a channel in the tree of section $S$, and a slice of thickness $dz$; the lunar gravitational force acting on the slice is,
\begin{equation}
F_{grav} \sim 0.566*10^{-6} m.s^{-2} * \rho dz dS,   
	\label{eq:02}
\end{equation}

with the dimension of the force. We will take $\rho = 10^{3} kg.m^{-3}$. In order to get an evaluation of the amount of pressure in the channel due to the gravitational force, we just write that the corresponding pressure gradient acting on the slice $dz$ is $S\dfrac{\partial p}{\partial z} dz$ is equal to the lunar gravitational force $F_{grav}$, 
\begin{equation}
S\dfrac{\partial p}{\partial z} dz = 0.566*10^{-6} m.^s{-2} * \rho S dz \Rightarrow \dfrac{\partial p}{\partial z} = 0.566*10^{-6} m.s^{-2}*\rho dz
\label{eq:03}
\end{equation}

Eventually, $\dfrac{\partial p }{dz} = 0.566*10^{3} Pa.m^{-1}$. Now, it remains to estimate the electric potential gradient or electric field, in volt (V) per meter, using the electrokinetic coefficient $C$, $\dfrac{d V}{dz} = C \dfrac{\partial P}{\partial z}$ (dimension $\dfrac{V.m^{-1}}{Pa.m^{-1}}$). In principle, $C$ is determined by laboratory experiments, 
\begin{equation}
C  = \dfrac{\varepsilon \zeta}{\eta \sigma_{f}}
\label{eq:04}
\end{equation}

$\sigma_f$ is the conductivity of the fluid (sap), $\varepsilon$ is the conductivity of the fluid (sap), $\zeta$ is the $\zeta$-potential (here between the fluid and the wood wall of the channel) and $\eta$ the viscosity of the fluid. In fact we know very little about the value of $C$ relevant to the trees. The value of the electric field can be of the order of 10 $mV/m$ only if $C$ is of the order of 10V/Pa. Such value is large compared to the values found in literature, generally relevant to rocks.

\section{\label{sec05} Discussion}
From a series of electrical SP data recorded on a Poplar tree in 2003 (see \cite{Gibert2006}), we propose to address the question of sap flow mechanism in trees. Today, for this mechanism, three sources are mentioned, which are not necessarily exclusive: capillarity, osmotic pressure, and evapotranspiration. We have discussed and critiqued the possibility of these mechanism in our introduction.\\ 

	In a recent study (see \cite{Courtillot2023}), we were able to measure the extent to which gravitational forces acting on Earth, through variations in its rotation speed and tilt, can influence the growth rates of tree rings in a Tibetan juniper forest for over 1000 years by modulating the incoming solar radiation. The growth rates of tree rings are linked to photosynthesis and, therefore, are also related to the flow of sap circulating in the xylem and phloem, which transport nutrients throughout the plant. At first order, like all fluids inside and on the surface of the Earth, the movement of this sap, regardless of its density, must be partially or entirely driven by lunar-solar tides, especially during the periods (<18.6 years) that are of particular interest in this paper. Gibert et al. (\cite{Courtillot2023}) emphasized that the sap flow and the recorded electrical signal were most likely linked by an  electrokinetic phenomena. Therefore, in order to explain the movement of sap from a physics perspective, we are dealing here with a simple harmonic pumping phenomenon (see \cite{Maineult2005,Maineult2008}). 

	As illustrated in Figure \ref{Fig:02}, the continuously recorded electrical signals in trees are non-stationary signals in the strict sense. This inherently prevents the direct use of a Fourier transform to determine the periodicities (\textit{i.e.}, lunar-solar tides) that may compose them. Furthermore, measurement vagary sometimes lead to interruptions in the recording system, resulting in discontinuous data. That is why we chose the Singular Spectrum Analysis (\textbf{SSA}) method to analyze the measurements of the Remungol poplar. This method helps limit possible errors in the interpretation of traditional spectra. \\
	
	In Figures \ref{Fig:05} and \ref{Fig:06}, we presented the five main pseudo-periodicities extracted from the electrical potential data of the poplar. All of them correspond very precisely to periods of lunar-solar tides (P$_1$, K$_1$, S$_1$, M$_f$, and K$_2$, \textit{cf.} our Table \ref{table01}). The periods that we have extracted from the electrical signals, which serve as proxies for sap flow, are within 0.01\% of the calculated lunar-solar tidal periods. This level of precision is remarkable and can typically only be achieved by expensive gravimeters costing tens of thousands of euros.
	
	The sum of the extracted components, which we believe to be tidal components, generally accounts for more than 70\% of the total variance in the recorded electrical signals.\\
	
	If we examine in more detail the signals recorded in different areas of the Poplar tree (root, branch, trunk) for a given tide, we observe that the waveforms associated with these tides are not modulated in the same way. 	These amplitude modulations can be completely distinct from each other, as is the case for the  tide and the electrodes E1, E8, E32, and E34 (cf. Figures 4). However, they can also be the same but phased in time. For example, we can observe that the patterns of the signals recorded by the electrodes E7 and E12 for the  tide are the same but shifted by approximately 2 weeks (see Figures \ref{Fig:05}). Undoubtedly, the differences in amplitudes and phases observed in the recorded signals of the Poplar tree are likely to be related to its physiological response. \\
	
	Based on these observations, we embarked on a mathematical resolution of the equations associated with the phenomenon that we believe to be responsible for the sap flow, aiming to evaluate the electrokinetic coefficient associated with the Poplar tree, similar to what is done in geophysics for rocks (see equation \ref{eq:04}). We obtained a value of 10 V/Pa, which is generally ten times higher than the values measured in rocks. However, it should be noted that trees are not rocks, as we can by the picture of the xylem geometry in Figure 01, sap  flow is not disrupted by any tortuosity, unlike in rocks. Furthermore, it is important to emphasize that sap is not water. \\
	
\newpage
\bibliographystyle{fchicago}
\bibliography{trees_arxiv.bib}

\section*{\label{app:A} Appendix A: The Singular Spectrum Analysis Method}
	The data, for which we have provided some examples (see Figures \ref{Fig:02} and \ref{Fig:02}), correspond to a class of signals that are non-stationary in the strict sense and piecewise continuous. We are not in optimal conditions for their analysis in the Fourier sense (see\cite{Claerbout1976,Kay1981}). This is why we opted for the Singular Spectrum Analysis (\textbf{SSA}), which has historically been developed, in the field of paleoclimatology, to analyze this type of signals (see \cite{Vautard1989,Vautard1992}). We will present a brief overview of this method, and all the calculation details can be found in the reference work by Golyandina et al. (\cite{Golyandina2018}).

\textbf{SSA} can be summarized in four steps. Let us consider a discrete time series ($\mathcal{X}_{N}$) of length N (N $>$ 2):

\begin{equation}
\mathcal{X}_{N} = (x_1,\ldots,x_N).
\end{equation}

\paragraph{Step 1: the Embedding Step\label{step01}}
$\mathcal{X}_{N}$ is divided into $K$ segments of length $L$ in order to build a matrix \textbf{X} with the dimensions $K \times N$ where $K= N-L+1$ will condition our decomposition. This is the first ''tuning knob''. Integrating \textbf{X} yields a Hankel matrix.   
\begin{equation}
\textbf{X} = 
\begin{pmatrix} 
x_{1}   & x_{2}  & x_{3} \cdots  & x_{K}\\
x_{2}   & x_{3}  & x_{4} \cdots  & x_{K+1}\\
x_{3}   & x_{4}  & x_{5} \cdots  & x_{K+2}\\
\vdots & \vdots & \vdots \ddots & \vdots\\ 
x_{L}   & x_{L+1}  & x_{L+2} \cdots  & x_{N}
\end{pmatrix}
\label{HankelMatrix}
\end{equation}

	For further details regarding the properties of square matrices with constant values along ascending diagonals, such as Hankel or Toeplitz matrices, we invite the reader to explore the work of Lemmerling and Van Huffel (\cite{Lemmerling2001}).

Embedding consists in projecting the one-dimensional time series in a multidimensional space of series $\mathcal{X}_{N}$ such that vectors $X_{i} = (x_{i}, \ldots,x_{i+L-1})^t$ belong to $\mathcal{R}^L$, where $K= N-L+1$. The parameter that controls the embedding is $L$, the size of the analyzing window, an integer between 2 and $N-1$. The Hankel matrix has a number of symmetry properties: its transpose $\textbf{X}^t$, called the trajectory matrix, has the dimension $K$. Embedding is a compulsory step in the analysis of nonlinear series. It consists formally in the empirical evaluation of all pairs of distances between two offset vectors, delayed (lagged) in order to calculate the correlation dimension of the series. This dimension is rather close to the fractal dimension of strange attractors that could generate that type of series. In this case, it is advised to select for the size of window $L$ very small values (and thus a very large $K$). On the contrary, for \textbf{SSA}, $L$ must be sufficiently large, so that each vector contains an important part of the information contained in the initial time series. 

\paragraph{Step 2: Decomposition in Singular Values\label{step02}}
Singular Value Decomposition (\textbf{SVD}, \textit{cf.} \cite{Golub1971}) of nonzero trajectory matrix \textbf{X} (dimensions $L \times K$ ) takes the following shape:
\begin{equation}
    \textbf{X} = \sum_{i=1}^d \sqrt{\lambda_i}U_i V_i^t
     \label{SVDform}   
\end{equation}

where the eigenvalues $\lambda_i (i=1, \ldots, L)$ of matrix $\textbf{S} = \textbf{XX}^T$ are arranged in order of decreasing amplitudes. Eigenvectors $U_i$ and $V_i$ are given by :
\begin{equation}
    V_i = \textbf{X}^T U_i/\sqrt{\lambda_i}
\end{equation}

The $V_i$ form an orthonormal basis and are arranged in the same order as the $\lambda_i$. Let $\textbf{X}_i$ be a part of matrix \textbf{X} such that:
\begin{equation}
     \textbf{X}_i=\sqrt{\lambda_i}U_i V_i^t.
     \label{SVDX}
\end{equation}

Embedding matrix \textbf{X} can then be represented as a simple linear sum of elementary matrices $\textbf{X}_i$. If all eigenvalues are equal to 1, then decomposition of \textbf{X} into a sum of unitary matrices is :
\begin{equation}
     \textbf{X} = \textbf{X}_1 +\textbf{X}_2 + \ldots + \textbf{X}_d
     \label{SumEleX}
\end{equation}

where $d$ is the rank of X ($d =  \textrm{rank}\  \textbf{X} = max\{i \vert \lambda_i > 0\}$). \textbf{SVD} allows one to write \textbf{X} as a sum of d unitary matrices, defined in a univocal way.

Let us now discuss the nature and the characteristics of the embedding matrix. Its rows and columns are sub-series of the original time series (or signal). The eigenvectors $U_i$ and $V_i$ have a time structure, and they can be considered as a representation of temporal data. Let \textbf{X} be a suite of $L$ lagged parts of ($\mathcal{X}$ and $X_1,\ldots, X_K$) the linear basis of its eigenvectors. If we let:
\begin{equation}
	 Z_i = \sum_{i=1}^d  \sqrt{\lambda_i}V_i
\end{equation}
with $i=1, \ldots, d$, then the relation (\ref{SVDX}) can be written:
\begin{equation}
	 \textbf{X} = \sum_{i=1}^d U_i Z_i^t
\end{equation}
that is, for the $j$th elementary matrix:
\begin{equation}
	 X_j = \sum_{i=1}^d z_{ji} U_i
\end{equation}
where $z_{ji}$ is a component of vector $Z_i$ . This means that vector $Z_i$  is composed of the ith components of vector  $X_j$.  In the same way, if we let:
\vspace{-6pt}
\begin{equation}
	 Y_i = \sum_{i=1}^d  \sqrt{\lambda_i}U_i
\end{equation}
we obtain for the transposed trajectory matrix:
\begin{equation}
     X_j^t = \sum_{i=1}^d  U_i Y_i^t
\end{equation}

which corresponds to a representation of the \textit{K} lagged vectors in the orthogonal basis ($V_1$, \ldots, $V_d$). One sees why \textbf{SVD} is a very good choice for the analysis of the embedding matrix, since it provides two different geometrical descriptions.

\paragraph{Step 3: Reconstruction\label{step03}}
As we have seen, $\textbf{X}_i$ matrices are unit matrices, and (as in the classical approach) one can ``re-group'' these matrices into a physically homogeneous quantity (or energetically homogeneous, etc…). This is the second ''tuning knob'' of \textbf{SSA}. In order to regroup the unit matrices, one divides the set of indices $i \{1, \ldots, d\}$ into m disjoint subsets of indices $\{I_1, \ldots, I_m\}$.

Let $I$ be the grouping of $p$ indices of $I=\{i_1,i_2, \ldots, i_p\}$; because (\ref{SumEleX}) is linear,  the resulting matrix $\textbf{X}_I$ that regroups indices I can be written:
\begin{equation}
    \textbf{X}_I = \textbf{X}_{I1} +  \textbf{X}_{I2}  + \ldots + \textbf{X}_{Im} 
    \label{SumElexGroup}
\end{equation}

This step is called regrouping the eigen triplets ($\lambda$, $U$ and $V$). In the limit case $m=d$, (\ref{SumElexGroup}) becomes exactly (\ref{SumEleX}), and we find again the unit matrices.

Next, how can one associate pairs of eigen triplets? This means separating the additive components of a time series. One must first consider the concept of separability.

Let $\mathcal{X}$ be the sum of two time series $\mathcal{X}^{(1)}$ and $\mathcal{X}^{(2)}$, such that $ x_i = x_i^{(1)} + x_i^{(2)}$ for any $i \in [1, N]$. Let $L$ be the analyzing window (with fixed length), and $X$, $X^{(1)}$ and $X^{(2)}$ the embedding matrices of series $\mathcal{X},\mathcal{X}^{(1)}$ and $\mathcal{X}^{(2)}$. These two sub-series are separable (even weakly) in Equation \eqref{SumEleX} if there is a collection of indices $\mathcal{I} \subset \{1, \ldots, d \}$ such that $\textbf{X}^{(1)} = \sum_{i\in \mathcal{I}} \textbf{X}_i$, respectively, if there is a collection of indices such that  $\textbf{X}^{(1)} = \sum_{i \not \in \mathcal{I}} \textbf{X}_i$. In the case when separability does exist, the contribution of $\textbf{X}^{(1)}$ for instance corresponds to the ratio of associated eigenvalues ($\sum_{i \in \mathcal{I}} \lambda_i$) to total eigenvalues ($\sum_{i=1}^d \lambda_i$).Still, in the case of relation (\ref{SumEleX}), let $\mathcal{I} = \mathcal{I}_1$ be the set of indices corresponding to the first time series, with corresponding matrix $X_{\mathcal{I}_1}$. If both this matrix and that corresponding to the second time series ($\textbf{X}_{\mathcal{I}_2}= \textbf{X} - \textbf{X}_{\mathcal{I}_1}$) are close or identical to a Hankel matrix, then the time series are approximately or perfectly separable. So, regrouping SVD components can be summarized by the decomposition into several elementary matrices, whose structure must be as close as possible to a Hankel matrix of the initial trajectory matrix (this is true on paper only; in reality things are much more difficult).

\paragraph{Step 4:the Diagonal Mean (Hankelization)\label{step04}}
The next, final step consists in going back to data space, that is, to calculate time series with length N associated with sub-matrices $\textbf{X}_I $. Let \textbf{Y} be a matrix with the dimensions $L \ast K$ and for each element $y_{i,j}$ we have $1 \leqslant i \leqslant L$ and $1 \leqslant j \leqslant K$. Let $L^*$  be the minimum and $K^*$  be the maximum. One always has $ N= L+K-1$. Finally, let $y^{\ast}_{ij} = y_{ij} \ \textrm{if} \  L < K \ \textrm{and} \  y^{\ast}_{ij} = y_{ji}$ otherwise. The diagonal average applied to the $k$th index of time series y associated with matrix \textbf{Y} gives:
\begin{equation}
y_k= \left\{ 
\begin{array}{cccc} 
\frac{1}{k} &\sum\limits_{m=1}^k y^{\ast}_{m,k-m+1} &\qquad 1 \leqslant k \leqslant L^\ast \\ 
\frac{1}{L^\ast} &\sum\limits_{m=1}^{L^\ast} y^{\ast}_{m,k-m+1} &\qquad L^\ast \leqslant k \leqslant K^\ast \\ 
\frac{1}{N-K+1} &\sum\limits_{m=k-K\ast +1}^{N-K\ast +1} y^{\ast}_{m,k-m+1} &\qquad K^\ast \leqslant k \leqslant N^\ast 
 \end{array}
 \right.
 \label{Hankelization}
\end{equation}

The relation (\ref{Hankelization}) corresponds to the mean of the element on the antidiagonal  $i+j = k+1$ of the matrix. For $k=1$, $y_1=y_{1,1}$. For $k=2$,  $y_2=(y_{1,2}+y_{2,1})/2$, {etc} \ldots 

Thus, one reconstructs the time series with length $N$ from the matrices of step \ref{step03}. If one applies the diagonal mean to unit matrices, then the series one obtains are called elementary series. Note that one can  naturally extend the \textbf{SSA} of real signals to complex signals. One only has to replace all transposed marks with complex conjugates. As mentioned above, step \ref{step03} is the most difficult part. 

	We have chosen one approach among many others: iterative \textbf{SSA}. Since the relation (\ref{SumEleX}) is linear, we can iterate the decomposition. We start with a small value of $L$ (we are looking for the longest period) that we increase until obtaining a quasi-Hankel matrix (step \ref{step01} and \ref{step02}). We then extract the corresponding lowest frequency component that it subtracted from the original signal. We increase again the value of $L$ to find the next component (shortest period). The algorithm stops when no pseudo-cycle can be detected or extracted. In this way, we scan the series from low to high frequencies.

\begin{figure}[H]
		\centering{\includegraphics[width=1.1\columnwidth]{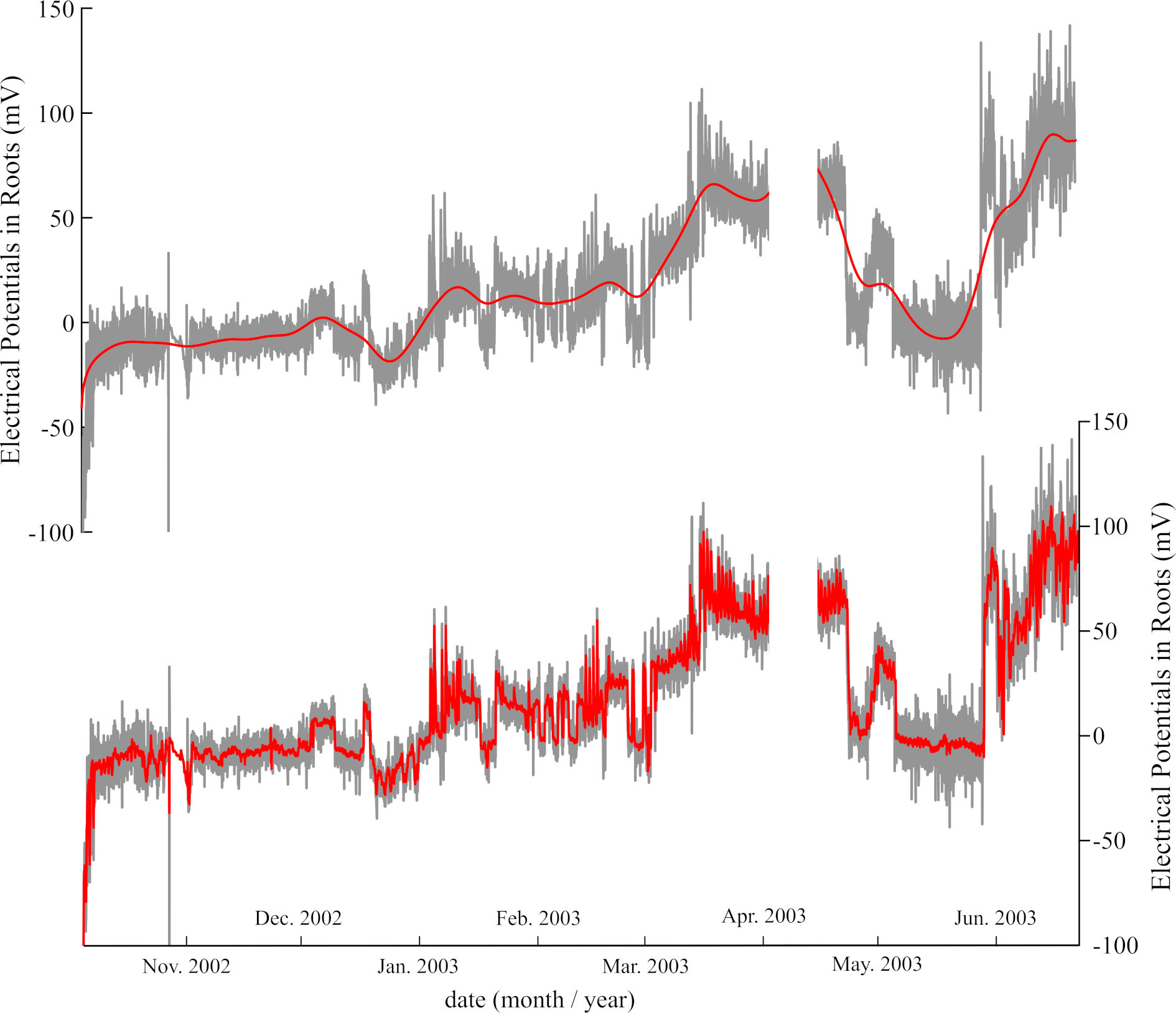}} 	
     	\caption{\redtext{At the top, in grey, the electrical signal measured on the poplar root by electrode E01. In red, the trend detected and extracted by the SSA method. Below, the same raw data in grey, and in red, the sum of the previous trend with the extracted pseudo-cycles, that will be presented in the next section.}}
		\label{Fig:08} 	
\end{figure}

As an example of the application of the SSA method, let us decompose the electrical signal from electrode E01 in the poplar root (see Figures \ref{Fig:01} and \ref{Fig:02}). As mentioned, the signal is discontinuous, with a gap in April 2003. Unlike traditional Fourier techniques, SSA allows the decomposition of any signal, continuous or not, stationary or not. In the upper part of Figure \ref{Fig:08}, we show the electrical potential in the root in grey and the trend extracted by SSA in red. In the lower part of the same Figure, we add to this trend all the cycles detected and extracted by SSA on the same electrode. These cycles and their spectra are listed and presented in Table \ref{table01} and Figures \ref{Fig:05} and \ref{Fig:06}.
\newpage

\end{document}